\documentclass[prc,aps,tightenlines,showpacs,twocolumn,floatfix,
unsortedaddress,superscriptaddress,preprintnumbers]{revtex4-1}

\usepackage{graphicx}
\usepackage{hyperref}
\usepackage[all]{hypcap}
\usepackage{amsmath,amssymb,amsbsy} 
\usepackage{color}
\usepackage[dvipsnames,usenames]{xcolor}

\hypersetup{colorlinks=true,citecolor=[rgb]{0,0,0.8},
linkcolor=[rgb]{0,0,0.8},urlcolor=[rgb]{0,0,0.8}}

\begin{document}
\preprint{LA-UR-17-31503, INT-PUB-17-056}

\title{Small bits of cold dense matter}
\author{S. Gandolfi}
\email{stefano@lanl.gov}
\author{J. Carlson}
\email{carlson@lanl.gov}
\author{A. Roggero}
\email{roggero@lanl.gov}
\affiliation{Theoretical Division, Los Alamos National Laboratory,
Los Alamos, New Mexico 87545, USA}
\author{J.~E.~Lynn}
\email{joel.lynn@gmail.com}
\affiliation{Institut f\"ur Kernphysik,
Technische Universit\"at Darmstadt, 64289 Darmstadt, Germany}
\affiliation{ExtreMe Matter Institute EMMI, GSI Helmholtzzentrum f\"ur
Schwerionenforschung GmbH, 64291 Darmstadt, Germany}
\author{S. Reddy}
\email{sareddy@uw.edu}
\affiliation{
Institute for Nuclear Theory, University of Washington, Seattle, WA 98195  \\
}
\begin{abstract}
   The behavior of QCD at high baryon density and
low temperature is crucial to understanding the properties of neutron stars
and gravitational waves emitted during their mergers.
In this paper we study small systems of baryons in periodic boundary conditions
to probe the properties of QCD at high baryon density.  By comparing
calculations based on nucleon degrees of freedom to simple quark models
we show that specific features of the nuclear spectrum, including
shell structure and nucleon pairing,  
emerge if nucleons are the primary degrees of freedom.  Very small systems
should also be amenable to studies in lattice QCD, unlike larger systems
where the fermion sign problem is much more severe.  
Through comparisons of lattice QCD and nuclear calculations it should
be possible to gain, at least at a semi-quantitative level,
more understanding of the cold dense equation of state
as probed in neutron stars.
\end{abstract}

\date{\today}


\maketitle

\section{Introduction}
Understanding quantitatively the properties of QCD at low temperature
and high baryon density remains one of the most challenging problems in
nuclear physics.
It is increasingly important as it governs the behavior of neutron stars 
including their mass-radius relations 
(see Refs.~\cite{watts:2016,Raithel:2017,Ozel:2016,Nattila:2017} and 
references therein), 
cooling, and the emission of 
gravitational waves in neutron star mergers~\cite{nstarmerger:2017}. 
At low to moderate 
densities it is natural to model neutron star matter via a 
system of interacting nucleons, and much progress has been 
made along these
lines~\cite{Gandolfi:2012,Holt:2013fwa,Gandolfi:2015,Hebeler:2015hla}
At very high densities the problem is also tractable: At 
low temperatures, the ground state will be a 
color-flavor-locked superfluid phase 
of quark matter with large superfluid pairing 
gaps~\cite{Alford:1999,Rajagopal:2001}.  At intermediate densities 
(several times nuclear saturation density) the situation is much less
clear.  One expects the 
dominant degrees of freedom to transition from nucleons to  quarks and gluons,
but the density at which this occurs remains very difficult to determine even qualitatively.

      Studying the neutron star mass-radius relationship and cooling of neutron stars 
has been an important goal to constraining the equation of state
through astrophysical observations~\cite{Steiner:2012}.  The recent observation of two 
solar mass neutron stars~\cite{Demorest:2010}, for example, has 
severely constrained possible equations of state at high density.  
Nevertheless these provide only some constraints on the equation of 
state, and less about the relevant degrees of freedom at high density.

      In this paper we study the behavior of small numbers of baryons 
at high density (or equivalently small volumes) in periodic 
boundary conditions.  It may soon be possible to study
these systems in lattice QCD, as the sign problem for small baryon number
is less severe. The sign problem grows exponentially with imaginary
time and is proportional to the number of nucleons times the mass of the
nucleon minus three halves the mass of the pion~\cite{Beane:2011}.
 We evaluate these systems in both nucleonic 
models and quark models, and identify specific features that 
arise in the spectra as the degrees of freedom change from nucleons to quarks.  
These features rely on the relatively high momenta 
and short distances that arise in small periodic volumes.  
The most important of these features are shell closures for small
numbers of nucleons
and pairing in open-shell systems. The latter is important even 
for very small systems, in particular for the $N=4$ systems we study.

The small lattice length $L$ needed to simulate high densities with a 
small number of baryons may be subject to significant corrections 
from long-range physics. However, we can expect that some of the 
longest-distance effects due to pions should be the same in 
nuclear and QCD simulations as long as nucleon-pion interactions 
are consistently included in the nuclear Hamiltonians. In any 
case the smallest box considered in this study (corresponding to 
$N=4$ at $\rho=0.48$ fm$^{-3}$) has a box size of $L\approx2.03$ fm.

It may also  be possible to perform lattice QCD and nuclear
simulations for unphysical heavy up- and down- quarks,
resulting in heavy pions. This would reduce the sign problem and 
the corrections due to
finite volume at the cost of unphysical pion masses.
Studying the transition even for 
high pion masses may be instructive as pion degrees of freedom may 
not play a very important role in the high-density phase transitions.
See Ref.~\cite{halqcd:2009} for some studies of $NN$ phase
shifts at high pion mass.

Small volumes will typically result in large excitation energies, reflecting
the wider spacing in the single-particle spectra, as discussed below.
This will be less true for comparison of different pairing symmetries
that are degenerate in the free-particle limit.   However, in general,
both the quantum Monte Carlo (QMC) and lattice QCD simulations will converge
more quickly with imaginary time for small systems.

Although studies of small systems are ill-suited to capture 
critical behavior and cannot precisely identify possible phase 
transitions, this work is well motivated because presently we 
lack even a qualitative understanding of how quark degrees 
of freedom might emerge at high density. The signatures we 
identify below suggest that these small systems could be quite valuable
in this regime.
Eventually one can add protons and/or hyperons to look at the 
density dependence of the symmetry energy and/or the presence of 
hyperons in neutron stars.  In these initial studies, 
we concentrate on pure neutron matter.

\subsection{Nucleonic Models:}  
       To study nucleonic matter we consider nonrelativistic nucleons
interacting via two-nucleon ($NN$) interactions:
\begin{equation}
H \ = \ -\sum_i \ \frac{\hbar^2 \nabla_i^2}{2 m} + \sum_{i<j} V_{ij},
\end{equation}
where the $NN$ interaction is taken as either the Argonne 
AV18~\cite{Wiringa:1995}, the
AV$8'$~\cite{Wiringa:2002},
or the local next-to-next-to-leading order (N$^2$LO) chiral interactions
of Ref.~\cite{Gezerlis:2014}.
At low densities such interactions should be able to faithfully 
reproduce the properties of QCD.
Of course the three- (and eventually four- and many-) nucleon forces 
will be present, and eventually play a significant role.  
The simple spectral features we identify below
will remain, though, even in a more sophisticated picture.  
These features depend primarily upon the single-particle 
states available to a nucleon in periodic boundary condition, 
indeed some are present even for noninteracting neutrons.
       
In this paper we consider only cubic simulation volumes with 
periodic boundary conditions. 
Other geometries, such as elongated volumes or different types of boundary 
conditions, might allow one to identify additional spectral
features~\cite{Bedaque:2004ax}, but since present lattice 
calculations typically 
use cubic symmetry, we will adopt it in this
study.
Below we show results 
for different densities with various numbers of neutrons in cubes of length $L$ on each side, and we show results as a function of the density $\rho = N/L^3$.

       The nuclear interaction models described above are defined in the continuum.  To maintain periodic boundary conditions, we add the contributions from periodic images for each pair:
\begin{equation}
    V_{ij} = \sum_{i_x,i_y,i_z=-M}^{M}  V\left[ r_{ij}+ L (i_x\
\hat{x} + i_y \  \hat{y} + i_z \  \hat{z})\right] \,,
\end{equation}
where $r_{ij}$ is the minimum separation in the periodic box and
$M = 1$ to $2$ images in each direction is sufficient to obtain 
periodic solutions since the $NN$ interaction is at most of pion
range.  We note that the longest-range corrections from pions ``wrapping
around'' the box are also present in lattice QCD
simulations~\cite{Colangelo:2003hf}.  For heavier pion masses
these long-range periodic images will play much less of a role.

For the larger number of neutrons ($N>4$), we restrict ourselves to solutions
fully symmetric under cubic rotations.  For the smallest systems, we also
investigate states that would correspond most closely to $p$-wave
($N=3,4$) or $d$-wave ($N=4$) solutions in the continuum.  The couplings to nonzero
angular momenta prove quite interesting in comparing neutron potential
models to quark models.

Calculations are performed using
QMC methods: Either the Green's function Monte Carlo or Auxiliary Field 
Diffusion Monte Carlo methods. More details are described in~\cite{Carlson:2015}.
The simulations are fairly simple as there are only a modest 
number of neutrons and the small volumes raise the energies 
of the excited states allowing for a more rapid convergence.

\subsection{Quark Model:}  

For comparison, we also consider very simple quark models of high-density QCD
in periodic boundary conditions.  These models are not intended to be 
realistic or predictive of the behavior of QCD in this regime, but they should 
illustrate possible alternative behaviors when deconfined quarks 
are the dominant degrees of freedom. 

We consider both free and interacting quarks in periodic boundary conditions.
In general the models can be written as:
\begin{equation}
H \ = \sum_i T_i + \sum_{i<j} V_{ij} \,.
\end{equation}
In the free case we consider for the kinetic energies $T_i$ of the  quarks
relativistic and nonrelativistic dispersion relations $T_i =  \sqrt{p_i^2 + m_i^2} $ and $T_i = m_i + p_i^2 / 2 m_i $, respectively. 
For the interacting quark model, we shall assume that chiral 
symmetry is not fully restored when quark degrees of freedom 
first manifest in the spectrum  and use a relatively large value 
of $m=300$~MeV. Under these conditions we treat
the quarks as nonrelativistic even for the small volumes that we consider.  

The pair potential is chosen in order to reproduce the pairing 
pattern expected from a Nambu-Jona-Lasinio-like
model~\cite{Klevansky:1992qe} where interactions are antisymmetric with
respect to color. 
Furthermore, for 3 colors and 2 flavors, we look for a color 
superconducting ground-state called the 2SC phase where two 
species are paired in a color and flavor anti-symmetric channel 
while the third does not interact with either and is effectively decoupled \cite{Alford:2007}. 
In the following we will consider the blue quarks to be the decoupled species.

For a given neutron number $N$ the occupation numbers in flavor-color-spin space are chosen in order to satisfy the following constraints
\begin{description}
\item[baryon number] $3N=(N_U+N_D)$
\item[color neutrality] $N_R=N_B=N_G$
\item[charge neutrality] $N_D=2N_U$ 
\item[spin neutrality] $N_\uparrow=N_\downarrow$ 
\end{description}

When quark degrees of freedom are manifest, it is appropriate to neglect 
confinement at the short-distance scales of relevance to our small volumes. 
We assume that the average confining interaction per baryon is a function of
the density only independent of baryon number at fixed density.
We then compare the evolution of the energy with baryon number 
at fixed density to the nucleonic models.

The shell structure can be influenced by interactions, though,
especially those that lead to pairing. 
To include this we shall consider a simple local potential of the form
\begin{equation}
\label{eq_pot}
V(i,j;\vec{r}_{i},\vec{r}_{j}) = \Lambda_{ij} V(r_{ij}) \,,
\end{equation}
where $i,j$ are multi-indices containing color, flavor and spin projection. 
In order to obtain the desired pairing structure expected in high 
density QCD,  we choose a simple short-range interaction 
\begin{equation}
V(r)=- \frac{4 \hbar^2}{\mu r_e^2} \alpha \beta^2 \frac{e^{-2\beta \frac{r}{r_e}}}{\left[1+\alpha e^{-2\beta\frac{r}{r_e}}\right]^2}.
\end{equation}
The matrix $\Lambda_{ij}$ is chosen to be anti-diagonal with entries equal 
to 1, 
$\mu$ is the reduced mass, and
\begin{equation}
\alpha=\sqrt{1-2\frac{r_e}{a_s}} \,, \quad\quad \beta = 1+\alpha \,,
\end{equation}
where $a_s$ and $r_e$ are the ($S$-wave) scattering-length and
effective-range respectively. 
In our 
calculations we choose $a_s=10$ fm 
and $r_e=0.1$ fm in order to be close to the unitary limit.

It is useful to compare relevant energy scales for 4 and 
14 neutrons in periodic boundary conditions.  In 
Table \ref{table:scale} we compare the lowest finite energy 
($k = 2 \pi /L$) free-particle modes in the box.
Since the quarks are light, it takes substantial 
energy to raise them to higher momentum states, 
as indicated in the table.  There are more degrees of 
freedom available, however, meaning that quarks will have a 
substantially lower Fermi energy for the case of 
weak interactions at high density.

\begin{table}
\begin{tabular}{lcccc}
\hline
Particle   & mass (GeV) &   N  & $\rho$ ($fm^{-3}$) & E (k=1, GeV) \\
\hline
Nucleon & 0.94 & 4 & 0.16 &  0.096 \\
rel q & 0.0 & 4 & 0.16 & 0.424 \\
rel q & 0.3 & 4 & 0.16 & 0.219 \\
non-rel q & 0.3 & 4 & 0.16 & 0.299 \\
\hline
Nucleon & 0.94 & 14 & 0.16 & 0.042 \\
rel q & 0.0 & 14 & 0.16 & 0.279 \\
rel q & 0.3 & 14 & 0.16 & 0.110 \\
non-rel q & 0.3 & 14 & 0.16 & 0.130 \\
\hline
Nucleon & 0.94 & 4 & 0.32 & 0.152 \\
rel q & 0.0 & 4 & 0.32 & 0.534 \\
rel q & 0.3 & 4 & 0.32 & 0.313 \\
non-rel q & 0.3 & 4 & 0.32 & 0.476 \\
\hline
\end{tabular}
\caption{Single particle energy levels for different baryon numbers
for SU(2) quarks and nucleons at different densities.
\label{table:scale}}
\end{table}

In very simple models the strength of the quark interaction is adjusted to reproduce the $N-\Delta$ mass splitting.  The total splitting is 320 MeV, which can be
compared to the single-particle energy splittings above.  
For example, for four neutrons, two nucleons have a momentum 
$|k| = 1$, while with quarks one can accommodate all quarks 
with $|k| = 0$ at the cost of twice the $N-\Delta$ mass 
splitting.  At low densities (large volumes) the four-neutron system would 
be preferred, but at high densities (small volumes)
having all the quarks at $|k| = 0$ would be preferable.

\section{Results for $\boldsymbol{N=3}$ and $\boldsymbol{4}$}

At present it is difficult to compute many-neutron states in lattice QCD because of the rapid growth in the number of correlators required and 
because the signal to noise ratio for small pion mass 
grows exponentially
with baryon number.  
Very small systems of 4 baryons may be easiest to simulate in
lattice QCD. For these systems we calculate states with different
quantum numbers as an additional probe of hadronic versus quark degrees
of freedom. 
We find distinctively different behavior even for
4 neutrons when comparing the  nuclear and quark models 
for different quantum states.  

Two neutrons in finite volume have been studied 
in lattice QCD (see for example Refs.~\cite{Beane:2006,Ishii:2007,Berkowitz:2015})  
and as two nucleons
using QMC methods in Ref.~\cite{Klos:2016}.
These results are directly tied to the phase shifts of the neutron-neutron
interaction via the L\"uscher formula. Systems of three and four neutrons
in external wells at low density have been studied in Ref.~\cite{Gandolfi:2017};
here we are interested in the behavior at high densities in periodic boundary
conditions to mimic lattice simulations.

\begin{figure}[htb]
\includegraphics[width=1.0\columnwidth]{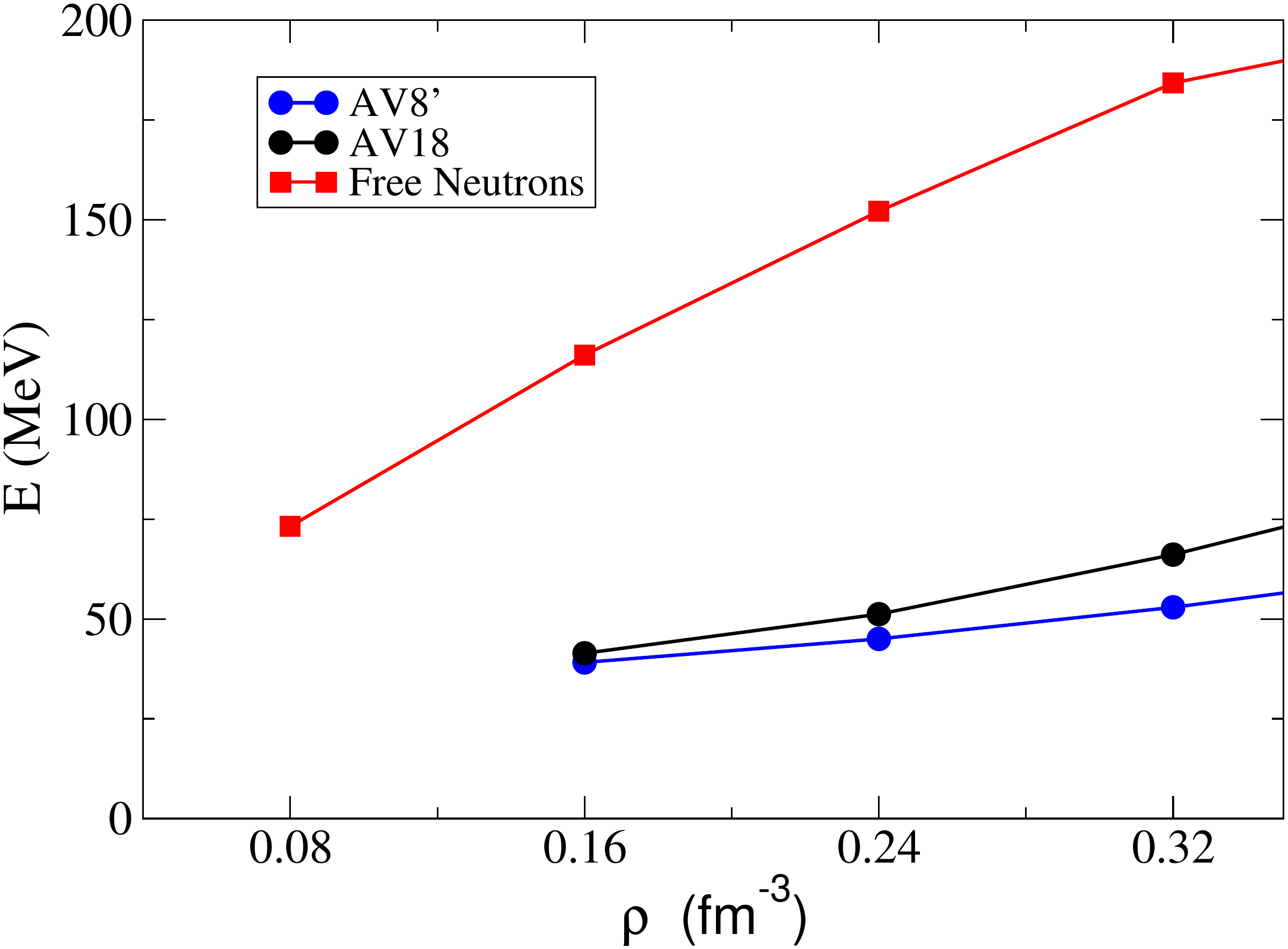}
\caption{Ground-state energies of 3 free neutrons and with
AV$8'$ and AV18 $NN$ interactions as a function of density.}
\label{fig:3n}
\end{figure}

For three neutrons we study low-lying states with the quantum numbers of
two neutrons with spin and total momentum zero, and with the extra neutron in a $|k|=1$ state. The spin of this unpaired neutron can be oriented
along or anti-aligned to the lattice equivalent of the angular momentum
giving something similar to $P_{3/2}$  or $P_{1/2}$ states.  As expected the
former are slightly lower in energy due to the spin-orbit splitting in the
neutron-neutron interaction. The total (including center-of-mass kinetic energy)
ground-state energies of three neutrons at different densities with the AV$8'$ and AV18 interactions are compared with free neutrons with the same boundary conditions in Fig. \ref{fig:3n}.

For four neutrons we study states with two neutrons paired to total
momentum and spin zero, and then either $s$-, $p$-, or $d$-wave pairing of the remaining
two dominantly $|k|=1$ neutrons.  The  $s$- and $d$-wave states have the spins coupled to
zero while the $p$-wave states must have the spins coupled to $1$ to maintain
antisymmetry.  The $s$-wave state is the same as that considered below
for up to 14 neutrons.

\begin{figure}[htb]
\includegraphics[width=1.0\columnwidth]{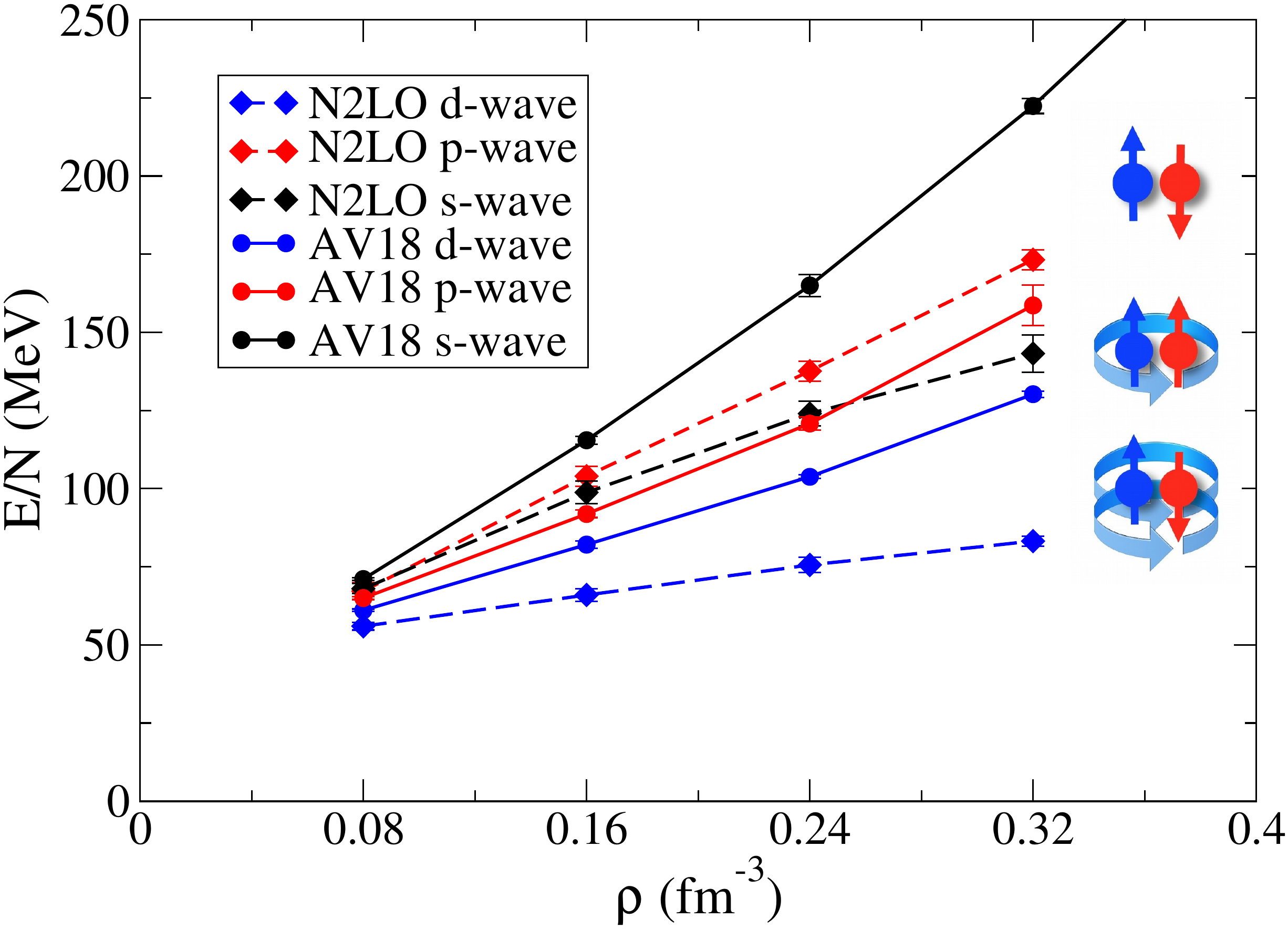}
\caption{Energy per neutron of 4 neutrons for states with different 
pairing symmetries as a function of density.}
\label{fig:4n}
\end{figure}

\begin{figure}[htb]
\includegraphics[width=1.0\columnwidth]{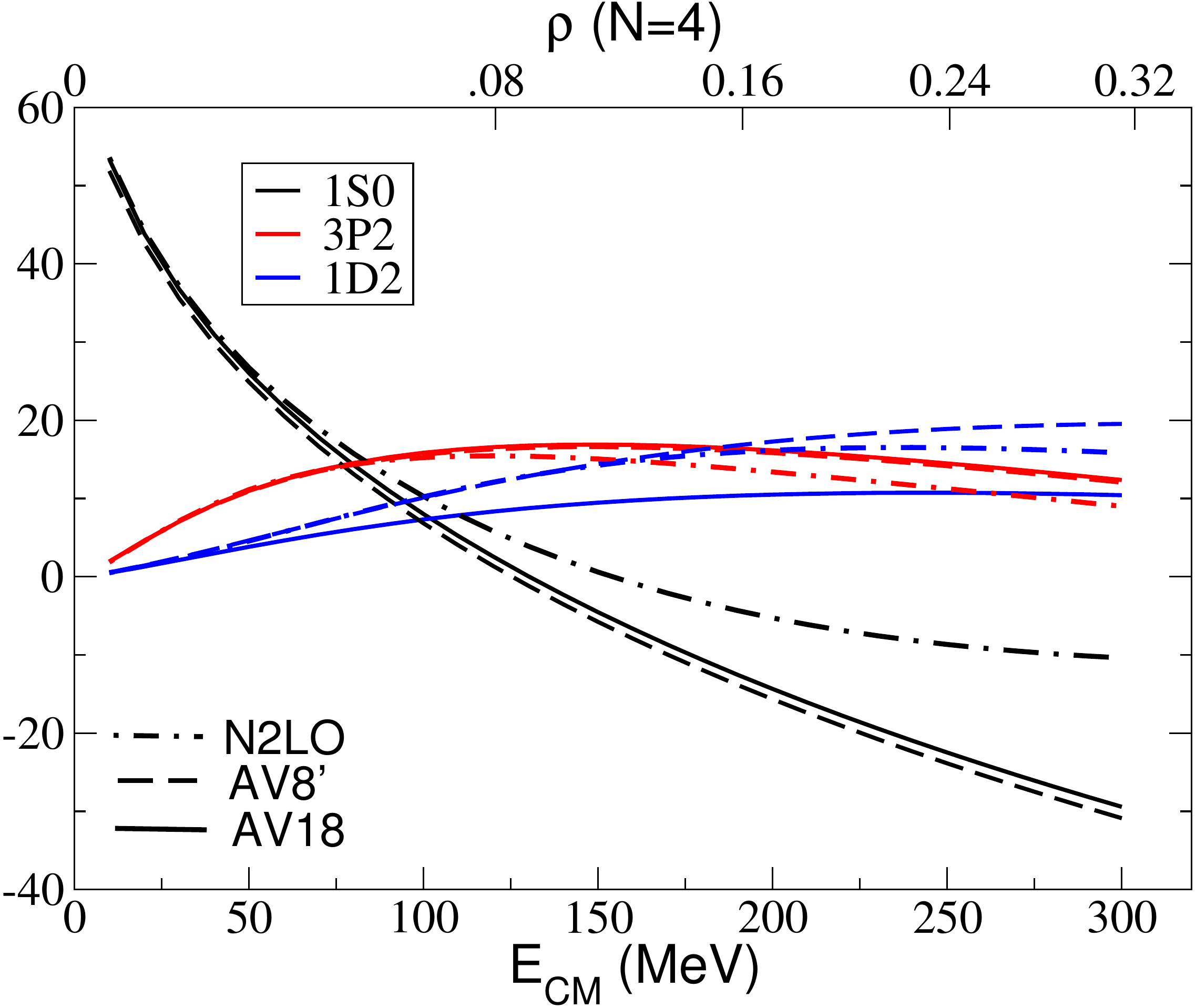}
\caption{$S$-, $P$-, and $D$-wave phase shifts for the 
different $NN$ interaction models.}
\label{fig:phases-4n}
\end{figure}

We find the initially surprising result (see Fig. \ref{fig:4n})
that neutron-neutron interactions
favor pairing in the $d$-wave state for small box sizes.  For free neutrons
the three different ($s$-, $p$- and $d$-wave) states 
would be degenerate.   The interaction between the
two neutrons in $|k|=1$ states coupled to zero total momentum dominates
the spectrum.  The relative momentum for this pair is quite large in these
small volumes, in the region of the repulsive $s$-wave neutron-neutron
interaction.  The relevant phase shifts are shown in Fig.~\ref{fig:phases-4n}.
At saturation density, two neutrons with momenta $+1$ and $-1$ are at a 
total energy of nearly 200 MeV, while at twice saturation
density the center-of-mass  energy is 300 MeV.
The strong $s$-wave repulsion disfavors the $s$-wave state, and the
periodic boundary
conditions favor the $L=2$ state, as the $d$-wave pairing is symmetric
across the periodic boundaries.
That is, a pair orbiting with $L=2$ feels an attractive interaction,
while for $L=1$, the periodic images interfere as the the relative coordinates
$r$ and $L-r$ and consequently $\vec{L} \cdot\vec{S}$ are oriented in opposite directions.

Results for the four-neutron calculations at different volumes are
shown in Fig.~\ref{fig:4n} for the AV18 and N$^2$LO $NN$ interactions. 
The different states all have very similar energies at half saturation density,
while the $d$-wave state is favored in all these models  
at $\rho = 0.16$ fm$^{-3}$ and above. At twice saturation
density the $s$-wave state is roughly 100 MeV higher than the $d$-wave state.

For free or paired quarks it is always advantageous to keep all the quarks
dominantly in $k=|0|$ states. The pairing energy that can be gained by promoting
some quarks to higher momentum is small compared to the energy cost
of promoting two or more quarks to $|k|=1$ states. These conclusions are unaltered by the addition of a gluon-exchange spin-interaction (which historically was invoked to explain the $N-\Delta$ mass difference) of the form \cite{Carlson:1983}
\begin{equation}
V_s(r_{ij})=\frac{2\alpha_s}{2m_im_j}\left[\frac{8\pi}{3}\vec{\sigma}_i\cdot\vec{\sigma}_j\delta^3(\vec{r}_{ij})+\frac{1}{r^3_{ij}}S^{(2)}_{ij}\right]\;,
\end{equation}
where $S^{(2)}_{ij}$ is the tensor operator. The energy difference between 
the $s$-wave and $d$-wave states favors the $s$-wave as the ground state,
since the $s$-wave is lower by $\Delta E$ per baryon $\approx 27$ MeV and $45$ MeV 
at 2 and 3 times nuclear saturation density, respectively.
This ordering is opposite to that seen in the neutron calculations.

\section{Results from $\boldsymbol{N=4}$ to 14 neutrons}

\begin{figure}[htb]
\includegraphics[width=1.0\columnwidth]{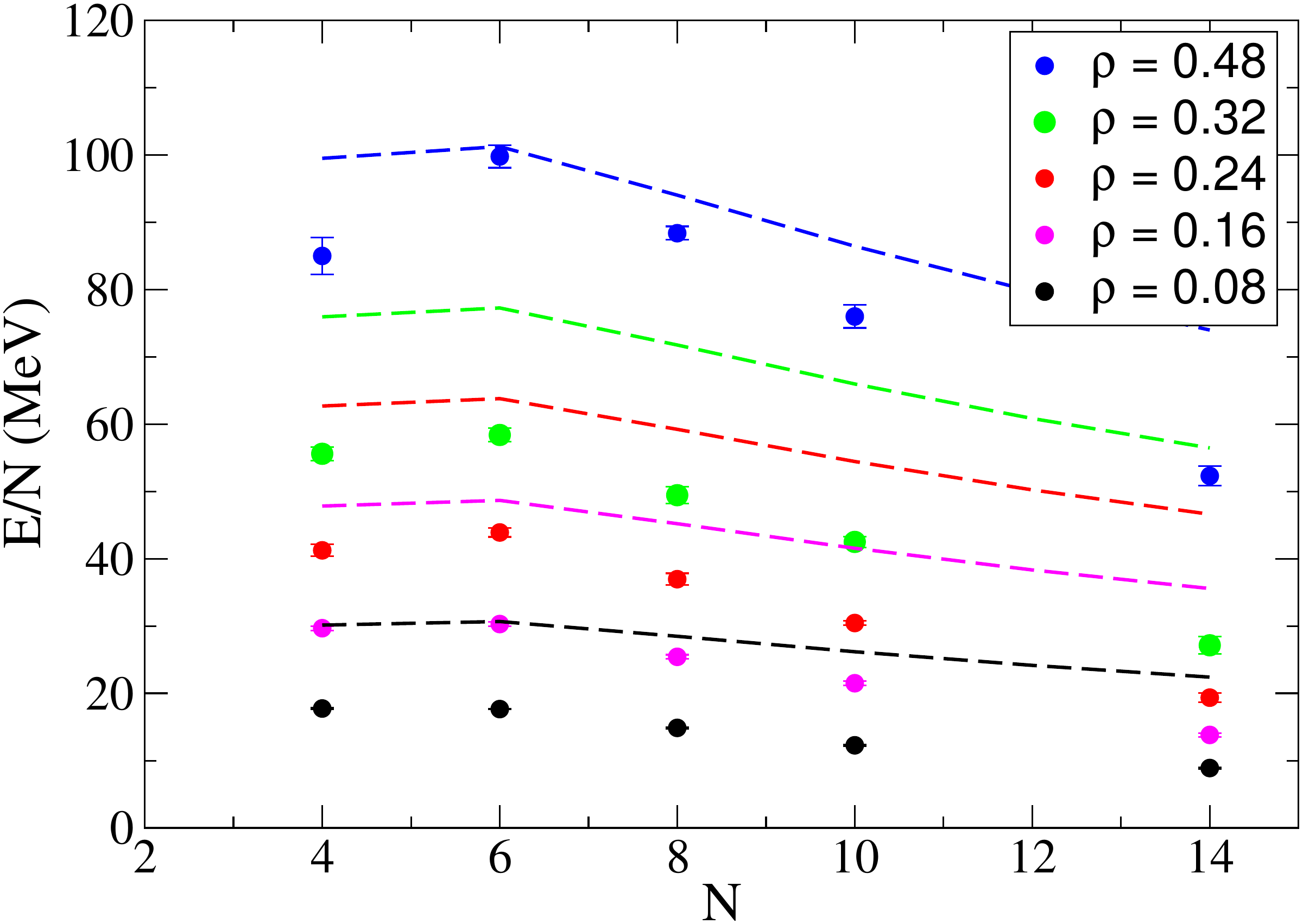}
\caption{Energies per neutron versus neutron number at different densities
for free neutrons (dashed lines) or with the AV18 $NN$ interaction (points). 
At lower densities the energies per neutron are quite small.  The upper
curves correspond to densities up to three times nuclear saturation
density, where the energies per particle (ignoring rest mass) are from
50-100 MeV per nucleon.  The minima at $N=14$ corresponds to a closed shell of neutrons with $|k| = 0,1$.}
\label{fig:shell}
\end{figure}

We also consider nuclear ground states 
from $N=4$ to 14 with even numbers of neutrons paired 
to spin zero with full cubic symmetry.  
In the continuum, this would correspond to an $s$-wave superfluid,  
which is expected to be the ground state at low 
densities~\cite{Gezerlis:2008,Gandolfi:2009b}.  
The special cases $N=2$ and 14 correspond to filled single-particle
nuclear shells and hence are expected to have lower 
energy per particle than the remaining (open-shell) systems.  
Similar behavior has been observed in calculations of neutrons
in external fields with either Woods-Saxon or harmonic 
wells~\cite{gandolfi:2011,Maris:2013}.
This is confirmed by our
numerical results as shown in Fig.~\ref{fig:shell}.  
The rest mass of the nucleons are not included in this figure.  

Note that the differences between different particle numbers are quite
significant at high density, of order 10 MeV per particle between adjacent $N$ and of
order 50 MeV per nucleon lower for $N=14$ compared to $N=6$. Such strong shell dependence, 
absent for free quarks, is also observed in the 
limit of strongly paired quark matter. 

\begin{figure}[htb]
\includegraphics[width=1.0\columnwidth]{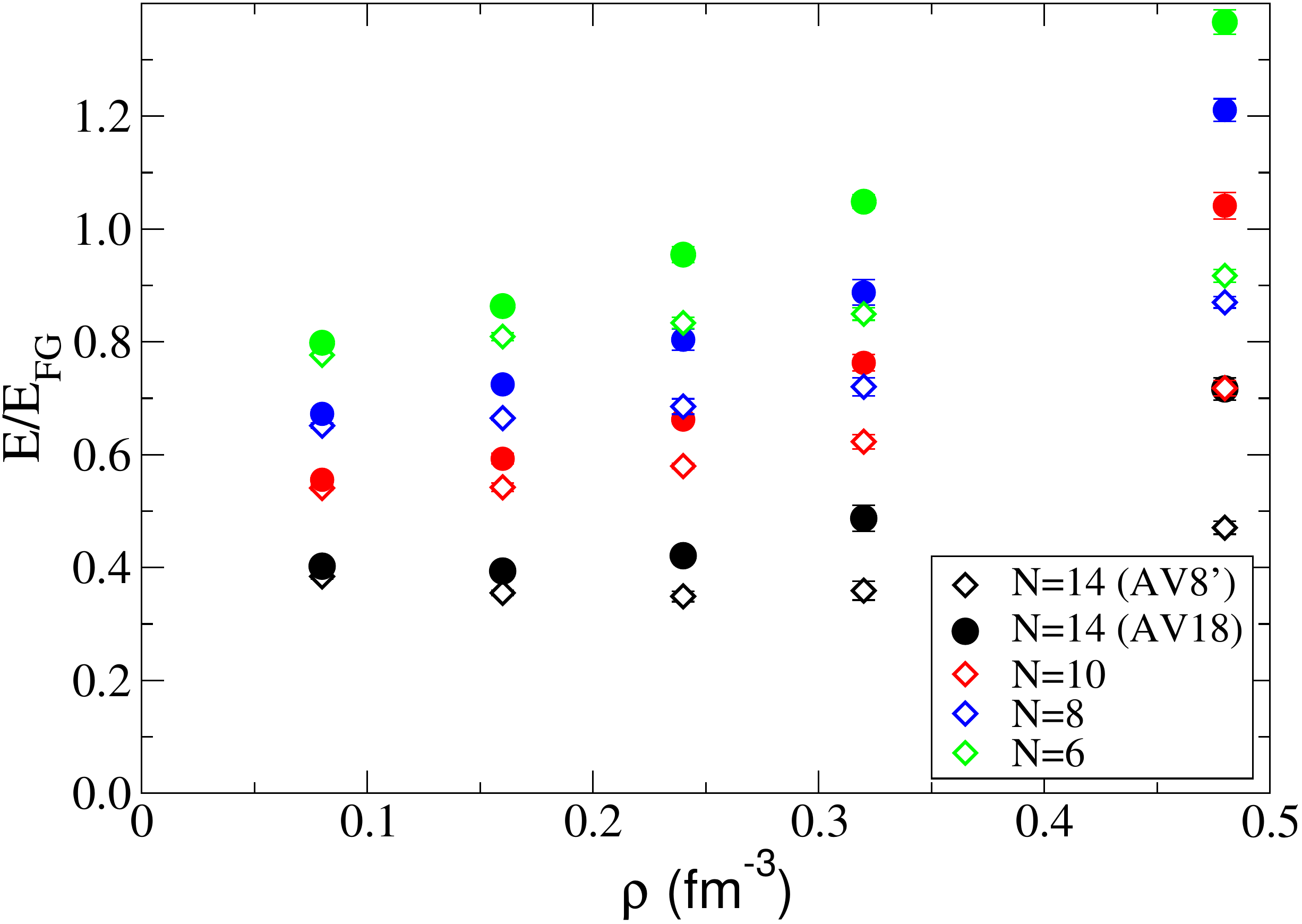}
\caption{Ratio of interacting to continuum Fermi Gas energies for the AV$8'$
(open symbols)  and AV18 $NN$ (closed symbols) interactions. }
\label{fig:eofg}
\end{figure}

The nuclear model dependence is fairly small at modest densities, but increases substantially at the highest densities considered. This is illustrated in Fig.~\ref{fig:eofg}, where we plot the ratio of energies to Fermi gas energies 
for different particle numbers and densities for both the AV$8'$ and AV18 interactions. The AV$8'$ model only fits the lower partial waves, and is therefore less reliable at moderate to high densities. 
The difference between AV18 and AV$8'$ is treated perturbatively in these
calculations.
The pattern of $E/N$ versus $N$ is the same for both interactions.

\begin{figure}[htb]
\includegraphics[width=1.0\columnwidth]{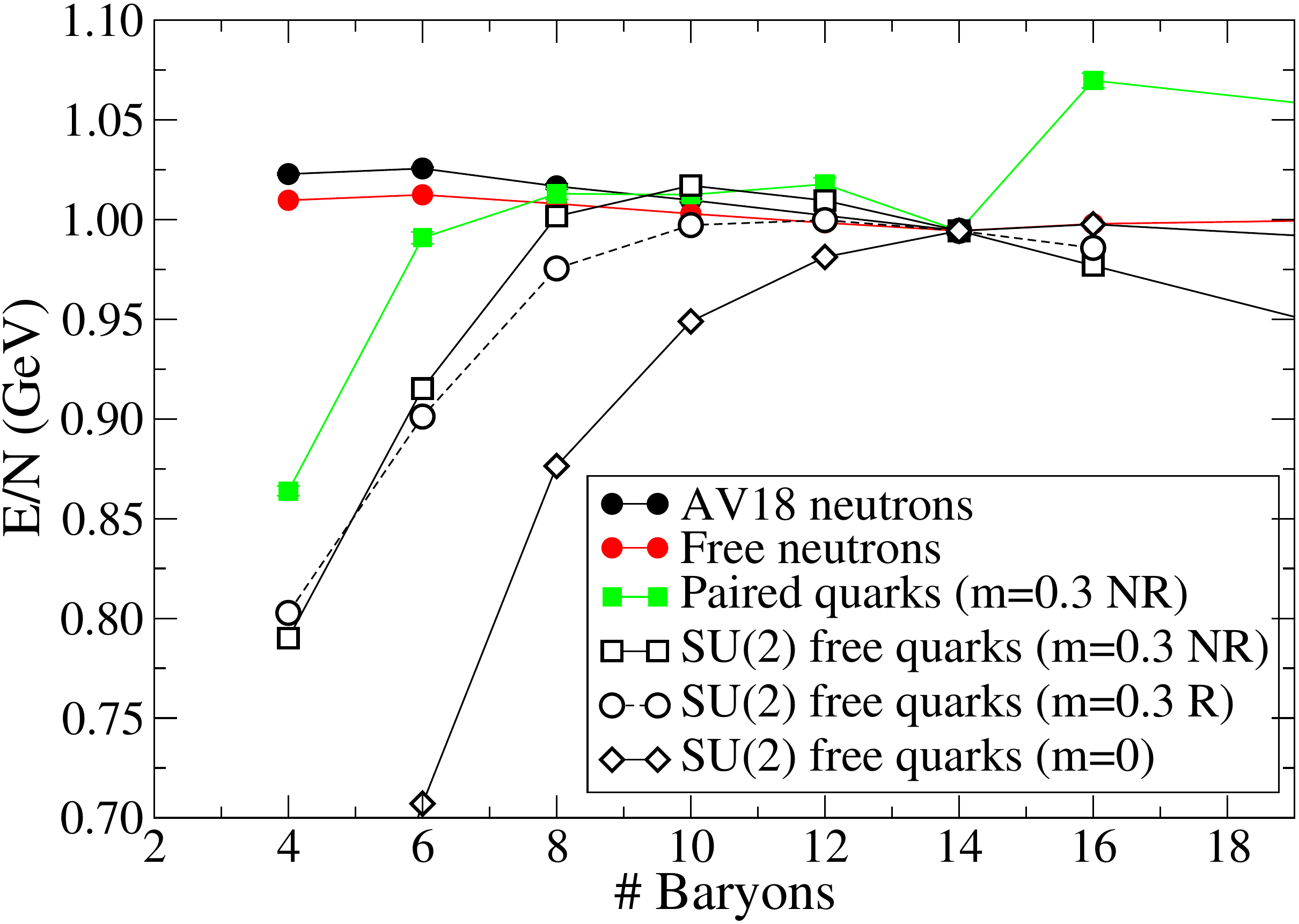}
\caption{Energy per neutron for nuclear and quark models at twice saturation density.  The behavior of the quark and nuclear models versus baryon number is quite different, particularly for small numbers.}
\label{fig:quarksandneutronsvsn}
\end{figure}

\begin{figure}[htb]
\includegraphics[width=1.0\columnwidth]{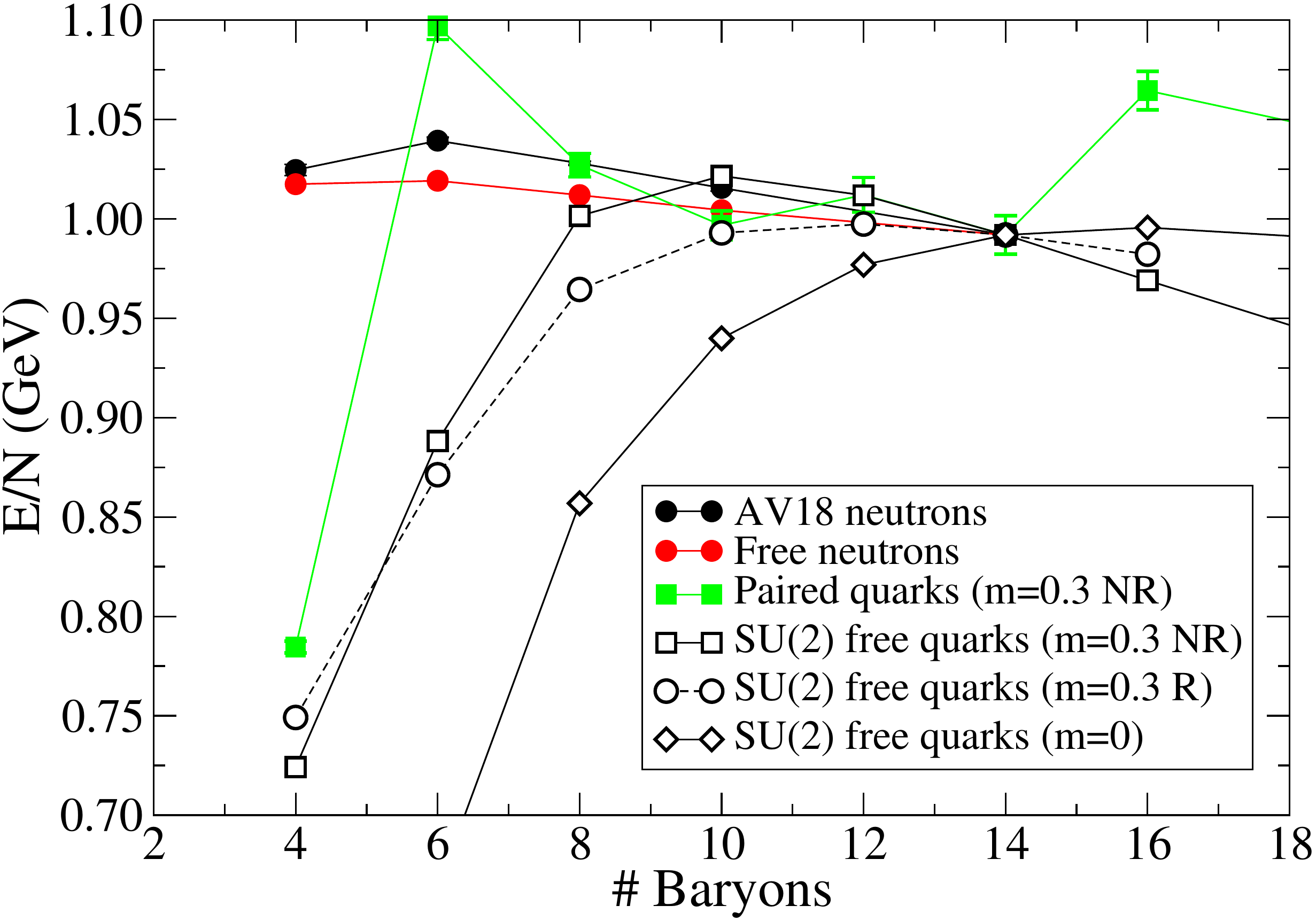}
\caption{Same as Fig.~\ref{fig:quarksandneutronsvsn} but at three times nuclear saturation density.}
\label{fig:quarksandneutronsvsn2}
\end{figure}

In Fig.~\ref{fig:quarksandneutronsvsn} we compare results of the nuclear
model with free neutrons and also free and paired quarks at twice
saturation density. The confinement energy in the quark model, which is
assumed to be constant with density, 
is adjusted to match the nuclear result for 14
interacting neutrons.
Note the dramatically different behavior versus baryon number in the
nuclear and quark models, particularly for small $N$. 
The same behavior is observed also at three times saturation density in 
Fig.~\ref{fig:quarksandneutronsvsn2}.

For both quarks and neutrons we find that the addition of interactions does 
not change the qualitative behavior apart from the cases $N=16$ and $N=6$ at high density. 
The sudden increase in energy for $N=16$ compared to $N=14$, which is 
caused by the filling of the $|k|=2$ momentum shell in the neutron case, 
has a similar nature for the quark model: At $N=16$, it is preferable for 
the (down) blue quarks to fill the $|k|=2$ shell instead of 
populating the (up) blue states as this will require breaking 
interactive pairs. The large energy at $N=6$ for the quark model 
compared to $N=4$ is due to the fact that for the latter all of the 
interacting quarks can reside in the $|k|=0$ shell while at $N=6$ 
two pairs of quarks have to fill the $|k|=1$ state. 
This large shell effect of $\approx 310$ MeV per 
baryon at $3$ times saturation density will necessarily indicate the 
dominance of quarks degrees of freedom. The feature is less pronounced
at twice saturation density where the, still large, gap is $\approx 120$ MeV.
This effect seems rather 
robust since even in the limit of noninteracting quarks the gap 
between $N_B=6$ and $N_B=4$ is still rather large, 
$\approx 170$ MeV for $\rho=0.48$ fm$^{-3}$ 
and  $\approx 120$ MeV for $\rho=0.32$ fm$^{-3}$.

\section{Conclusions}

In this work we have studied the behavior of the energy 
per baryon in high density neutron matter for small volumes.
These types of simulations can be valuable in understanding the
relevant degrees of freedom at high baryon chemical potentials, 
in particular to begin to identify the regime where nucleon degrees
of freedom dominate and where full QCD simulations of quarks
and gluons are required.

While these simulation volumes are small, particularly for $N=4$, and may
suffer from important finite-size effects beyond the periodic pion 
potentials discussed here, further studies with larger quark masses
and the $NN$ potentials derived from them are warranted.
Eventually lattice QCD studies with larger $N$ at physical pion masses
will be important.

Even for very small systems, we find a particular 
pairing pattern in neutron simulations.
For four neutrons the $d$-wave state is favored at 
high density for the two neutrons in the $|k|=1$ state,
due to a combination of the $s$- and $d$-wave phase shifts 
and the periodic boundary conditions. In contrast, for quark 
models the $s$-wave pairing is always favored 
to avoid filling the $|k|=1$ states.

A clear signature of the dominance of strongly paired quarks as
degrees of freedom can also be seen in the large energy gap 
between $N_B=4$ and $N_B=6$, in particular for paired quarks 
compared to nucleons at 2 (or 3) times saturation density. 
For the spatially symmetric states,
we expect an increase of $\approx 120$ MeV per baryon 
($\approx 310$ MeV per baryon) to be compared with a much 
smaller gap of $\approx 4$ MeV per baryon 
($\approx 15$ MeV per baryon) for the interacting nuclear case. 

Simulations of three- and four-nucleons ($A=3$ and 4 nuclei) 
have already been completed
with higher quark masses, so it is reasonable to expect that similar
comparisons of neutron simulations to full lattice QCD 
could be completed in the near future. A direct comparison would require
reasonable extractions of the $NN$ scattering phase shifts for these
quark masses.  Such a direct comparison could have 
dramatic implications for the behavior of QCD at low temperatures and
high baryon density and hence the mass-radius relation of neutron stars
and the gravitational waves emitted during their mergers.

\emph{Acknowledgments:}
The work of J.C. and S.G. was supported by the NUCLEI SciDAC program,
by the U.S. DOE under contract DE-AC52-06NA25396, and by the LANL LDRD
program. A.R. was supported by NSF Grant No. AST-1333607 and 
by DOE Grant No. DE-AC52-06NA25396.
The work of J.L. was supported by the ERC Grant No. 307986 STRONGINT and
the BMBF under Contract No. 05P15RDFN1.
The work of S.R. was supported by DOE Grant No. DE-FG02-00ER41132.
Computational resources have been provided by Los Alamos Open
Supercomputing.
We also used resources provided by NERSC, which is supported by the US
DOE under Contract DE-AC02-05CH11231.

\end{document}